\documentclass[10pt,draft]{dis03}
\usepackage{epsf,amsmath}

\begin{document}

\title{Transversity and  single spin asymmetries
}

\author{O.V. Teryaev, \\
         Bogoliubov Laboratory of Theoretical Physics, JINR, \\ 141980 Dubna, Russia}

\maketitle

\begin{abstract}
\noindent 
The transverse spin of interacting light quarks is described by
two independent non-perturbative inputs, twist-2 transversity  
and twist-3 quark-gluon correlators, while for free massive quarks 
they are related by the equations of motion. Similar relation emerges also 
in a model, where quark-hadron transition is treated in a probabilistic way.
The single spin asymmerty in SIDIS appears due to the interplay of 
transverity with Collins fragmentation function, which appears to be  
of twist 3, when defined in a Lorentz invariant way. 

\end{abstract}

\section{Transverse spin for free and interacting quarks} 

The basic property of the quantum theory of angular momentum is 
its transformation for the rotation from the longitudinal 
to transverse direction, making the interference helicity-flip
effects the probabilistic ones. The simplest way to see that 
is to observe that $\sigma_z$ Pauli matrix is a diagonal one,
contrary to $\sigma_x$ and $\sigma_y$, and this is a core 
of the role played by the transversity basis. 
The important further distinction comes when free and interacting 
particles are considered. The state of free particle
is completely described by its wave function, which determines 
the density matrix $\rho = 1/2(\hat p + m)(1+ \hat s \gamma_5) $, and the coefficients of 
all mutually orthogonal Dirac matrices are fixed. At the same time, 
the parametrization of transverse polarized nucleon matrix element \ref{ET82} 
of light-cone 
quark operator leads to the similar 
expression, which may be therefore interpreted as a quark density 
matrix
\begin{equation}
q(x) \hat P + M g_T(x) \hat S_T \gamma_5 +h_1 (x) \hat P \hat S_T \gamma_5
\end{equation}
The mass of free particle is now substituted by a parameter of the order of hadron 
mass \cite{ET82}. 
The coefficients of various Dirac matrices are 
now different parton distributions. The transverse spin may
be therefore described by the axial or pseudotensor terms.
As the first one enters together with hadron mass,
it corresponds to the subleading twist 3, while 
the second is just the twist 2 transversity. 

The independence of these two functions is not a complete one.
The operator equations of motions \ref{ET82}, which are crucial to 
guarantee both electromagnetic and colour gauge invariance 
lead to the integral relations between them,
including also another important non-perturbative 
input, namely the quark-gluon correlator $b$: 
\begin{equation}
M g_T(x) +M \int dy b(x,y))= m h_1 (x)
\end{equation}
 There are two extreme cases: when one may neglect 
these correlators, and the quark mass $m$ is of order 
of the hadron one $M$, one is coming back to the 
free particle case and transversity happens to be 
equal to other spin distributions \ref{ET82}.

In the opposite case of very light quark, realized in Nature, 
transversity decouples, so that the very existence
of such independent distribution is closely related to 
the spontaneous chiral symmetry breaking,
being the mechanism of generation of the most of the observable mass.

The relation between various spin distributions is restored 
in some models. Let me briefly discuss
the recent calculation \cite {ETZ} of transversity, which uses the  
elegant interpretation \cite{GJJ} as a DIS for the interfering
vector and scalar currents. 
The obtained result is convoluted with the probabilistic 
momentum space distribution 
of quark in nucleon. 
This allows to perform the calculation in complete analogy with 
the earlier treatment \cite{PZ} of unpolarized and polarized quark distribution.   
The obtained result looks especially simple when one neglects the quark mass.
In the same approximation within this model \cite{PZ} one reproduces the standard 
Wandzura-Wilczek (WW)
relation, so it may be called the WW approximation for transversity:
\begin{equation}
g_T(x)+g_1(x)= h_1 (x).
\end{equation}
In particular, taking into account the Burkhardt-Cottingham sum rule,  
 \begin{equation}
2 \int dx g_1(x)= \int dx h_1 (x),
\end{equation}
which is not far from the model estimation \cite{AV03}.
Although model, in principle, may be applied to any quark flavour,
one may have the violation of Soffer inequality if $g_1(x)< 0$, i.e. for d-quarks.
This shows that the adopted probabilistic consideration of parton-hadron transition (while
the interference effects are taken into account on the level of vector and scalar currents only)
works worse for $d$-quarks. One may recall, that standard WW approximation for d-quarks 
may also violate the positivity bound \cite{ST}.  

\section{T-odd fragmentation and distribution functions: twist and universality}

The appearance of transversity in SIDIS may be only in the combination with chiral-odd
fragmentation function, the most popular one being the T-odd Collins fragmentation function \cite{AV03,metz},
  
As this quantity does not correspond to the standard one, being the intrinsic transverse
momentum dependent, the notion of twist is not so trivial. 
The relevant tensor structure is $M^{-1} H(z,k_T)\epsilon_{\mu \nu P k_T}$, 
factor $1/M$ appearing due to the dimensional 
reasons. As soon as $M$ and $k_T$ are of the same order, any power suppression is absent 
which is usually considered as a manifestation of leading twist (two) contribution. 

At the same time, the consideration of transverse momentum dependent functions did not yet
receive the full field-theoretical justification beyond the Born approximation. 
The exception is provided by their 
expansion over the powers of $k_T$. In particular, linear in $k_T$ terms constitute 
nothing else but WW approximation to inclusive \cite{hera96}
and exclusive \cite{n} processes.  

It is interesting, that WW approximation may be also obtained in the coordinate representation
\cite{BB}, when no reference on the transverse direction is needed. 
This suggests the possibility \cite{OT03} 
to consider the Collins function in the coordinate space as well. 
The tensor structure is than  $ M I(z)\epsilon_{\mu \nu P z}$, so that the transverse 
momentum of the quark $k_T$ is not present and the factor $M$ appears now in the numerator, 
rather in the denominator, due to the dimensional reasons.  
However, when one is calculating the weighted (with $p_T$, being the transverse momentum of the produces {\it
pion}) average, 
one is coming to the same result, and $I(z)$ corresponds to the {\it moment} of the Collins 
function: 
\begin{equation}
I(z))  \sim \int d k_T^2  \frac {k_T^2 }{M^2} H_1(z,k_T^2 ),
\end{equation}
and the factor $M^2$ in the denominator of the r.h.s. is exactly the 
one resulting from the various appearance of $M$ in the definitions of $H$
and $I$.
The proportionality of $I$ to $M$ makes this function the twist-3 one,
according to the standard classification (like,say, the functions $g_T$ and $g_2$).
The trace of the twist-3 nature of $I$ is the necessity of $1/M$ factor 
in the definition of the weighted cross-section, 
\begin{equation}
\int d p_T^2  \frac {p_T }{M} \frac{d \Delta \sigma}{d p_T^2},
\end{equation}
so that the $M$ should compensate the dimension of $P_T$, in order to 
get the unsuppressed expression. Consequently, if such a factor is not introduced
by hand, the result should be suppressed like $M$, analogously to any other 
twist-3 observable. This suppression is absent, when 
$P_T$ of order $M$ (where factorization is, generally speaking, unapplicable beyond 
Born approximation) are considered.
Let us also briefly comment on the complementary mechanism of generation of Single Spin Asymmetry
in SIDIS, namely the Sivers function. 
As it was already discussed earlier \cite{OT00}, such a function can be only {\it effective},
(such a nortion first suggested in \cite{BMP})
or non-universal (like it is referred to now \cite{metz}),  
in the sense, that the imaginary phase emerges in the interaction, involving
also the hard scattering and depending on its type. 
In other words, the respective cut, providing the imaginary phase, 
involves both hard and soft variables.
Moreover \cite{OT00}, the model calculation \cite{BHS} of Brodsky, Hwang and Schmidt,
clearly exhibits this property, as well as elegant interpretation of Collins \cite{col02}. 
Let us also mention in this connection, that the similar calculation was performed
earlier in twist-3 QCD \cite{KT} for the crossing related process of 
dilepton photoproduction. That result, when continued to the region $P_T \sim M$,
will also not have any power suppression and looks formally as a twist 2 one.  

At the same time, the imaginary phase in the Collins function should come from the 
cut with respect to the jet mass, and is therefore universal. 
The recent analysis \cite{metz} seems to confirm this picture.

\section{Conslusions}

The description of Collins function in coordinate space leads to its twist being equal to 3,
while the absence of power suppression is related to the definition of observables. 
The Sivers function is expected to be an effective, or non-universal one,
contrary to Collins function.  
The earlier calculation in twist 3 QCD \cite{KT}, when continued to 
low $P_T$ region, shows no power suppression and may be considered as an estimate of effective
Sivers function.



\begin{thebibliography}{0}

\bibitem{ET82}
A.~V.~Efremov and O.~V.~Teryaev,
Sov.\ J.\ Nucl.\ Phys.\  {\bf 36}, 140 (1982)
[Yad.\ Fiz.\  {\bf 36}, 242 (1982)].

\bibitem{ETZ}
A.V. Efremov, O.V. Teryaev and P. Zavada, work in progress.




\bibitem{GJJ}
G.~R.~Goldstein, R.~L.~Jaffe and X.~D.~Ji,
Phys.\ Rev.\ D {\bf 52}, 5006 (1995);
B.~L.~Ioffe and A.~Khodjamirian,
Phys.\ Rev.\ D {\bf 51}, 3373 (1995)


\bibitem{PZ}
P.~Zavada,
Phys.\ Rev.\ D {\bf 67}, 014019 (2003) 

\bibitem{AV03} A.V. Efremov, these Proceedings. 


\bibitem{ST}
J.~Soffer and O.~V.~Teryaev,
Phys.\ Lett.\ B {\bf 490}, 106 (2000)


\bibitem{metz} A. Metz, these Proceedings.

\bibitem{hera96} O.~V.~Teryaev,
arXiv:hep-ph/0102296.

\bibitem{n}
I.~V.~Anikin and O.~V.~Teryaev,
Phys.\ Lett.\ B {\bf 509}, 95 (2001)
[arXiv:hep-ph/0102209].

\bibitem{BB}
P.~Ball and V.~M.~Braun,
Phys.\ Rev.\ D {\bf 54}, 2182 (1996)
[arXiv:hep-ph/9602323].

\bibitem{OT03} O.V. Teryaev, work in progress.

\bibitem{OT00}
O.~V.~Teryaev,
RIKEN Rev.\  {\bf 28}, 101 (2000);
Nucl.\ Phys.\ A {\bf 711}, 93 (2002).
Czech.\ J.\ Phys.\  {\bf 53}, 47 (2003).

\bibitem{BMP} D.~Boer, P.~J.~Mulders and O.~V.~Teryaev,
Phys.\ Rev.\ D {\bf 57}, 3057 (1998)
[arXiv:hep-ph/9710223].

\bibitem{BHS}
S.~J.~Brodsky, D.~S.~Hwang and I.~Schmidt,
Phys.\ Lett.\ B {\bf 530}, 99 (2002)

\bibitem{col02}
J.~C.~Collins,
Phys.\ Lett.\ B {\bf 536}, 43 (2002)



\bibitem{KT}
V.~M.~Korotkiian and O.~V.~Teryaev,
Phys.\ Rev.\ D {\bf 52}, 4775 (1995).









\end{thebibliography}
\end{document}